\documentstyle[preprint,aps]{revtex}
\tighten
\begin{document}
\draft
\title{Crossover effects in a discrete deposition model
with Kardar-Parisi-Zhang scaling}
\author{Anna Chame and F. D. A. Aar\~ao Reis}
\address{
Instituto de F\'\i sica, Universidade Federal Fluminense,\\
Avenida Litor\^anea s/n, 24210-340 Niter\'oi RJ, Brazil}
\date{\today}
\maketitle

\begin{abstract}

We simulated a growth model in $1+1$ dimensions in which  particles are
aggregated according to   the rules of  ballistic deposition with
probability $p$ or  according to the rules of random deposition with 
surface relaxation (Family model) with  probability $1-p$. For any
$p > 0$, this system is in the Kardar-Parisi-Zhang ($KPZ$) universality class,
but it presents a slow crossover from the Edwards-Wilkinson class ($EW$) for
small $p$. From the scaling of the growth velocity, the parameter
$p$ is connected to the coefficient of the nonlinear term of the $KPZ$
equation, $\lambda$, giving $\lambda \sim p^\gamma$, with $\gamma =
2.1\pm 0.2$. Our numerical results confirm the interface width scaling in the
growth regime as $W\sim \lambda^\beta t^\beta$ and the scaling of the
saturation time as $\tau\sim \lambda^{-1} L^z$, with the expected exponents
$\beta =1/3$ and $z=3/2$ and strong corrections to scaling for small
$\lambda$. This picture is consistent with a crossover time from $EW$ to $KPZ$
growth in the form $t_c\sim \lambda^{-4}\sim p^{-8}$, in agreement with
scaling theories and renormalization group analysis. Some consequences of the
slow crossover in this problem are discussed and may help investigations of
more complex models.

\end{abstract}
\date{\today}

\widetext
%\narrowtext

\section{Introduction}
\label{intro}

Surface growth processes and deposition of thin films are of
great interest due to potential technological applications (such as production
of nanostructures for microelectronic devices, the possibility of growth of
designed quantum objects, magnetic storage devices, among others) and due to
the fundamental role these systems play in  non-equilibrium statistical physics
\cite{Stanley,Villain}. Several models have been investigated in the last
decade, most of them involving one kind of particle and a simple
microscopic aggregation rule. The competition between different
growth mechanisms have received less attention, but
is essential to describe some practical situations, such as growth
of materials designed to have specific electronic, mechanical or magnetic
properties, which involves deposition of two or more chemical species.
In this framework, some authors considered growth models with
two kinds of particles and different aggregation
rules\cite{Cerdeira,poisoning,kotrla}. Other situations involving competition
between two growth mechanisms have also been considered
\cite{Albano1,Albano2,tales}. 

These models usually show crossover effects from one dynamics at small
times $t$ or short length scales $L$ to another dynamics at long $t$ and
large $L$. One typical example is Kardar-Parisi-Zhang ($KPZ$) growth
at small nonlinearities~\cite{KPZ}. The Langevin-type equation
\begin{equation}
{{\partial h}\over{\partial t}} = \nu{\nabla}^2 h + {\lambda\over 2}
{\left( \nabla h\right) }^2 + \eta (\vec{x},t) ,
\label{kpz}
\end{equation}
known as $KPZ$ equation, was proposed as a hydrodynamic description of kinetic
surface roughening. Here $h$ is the height at the position $\vec{x}$ in a
$d$-dimensional substrate at time $t$, $\nu$ represents a surface tension,
$\lambda$ represents the excess velocity and $\eta$ is a Gaussian
noise~\cite{Stanley,KPZ} with zero mean and variance $\langle
\eta\left(\vec{x},t\right) \eta\left(\vec{x'},t'\right) \rangle = D\delta^d
\left(\vec{x}-\vec{x'}\right) \delta\left( t-t'\right)$. When the coefficient
$\lambda$ of the nonlinear term is small,  a crossover is observed from linear
growth ($\lambda = 0$, known as Edwards-Wilkinson theory - $EW$)~\cite{EW} to
$KPZ$ behavior.

In discrete models, the interface width, which characterizes the roughness of
the interface, is defined as
\begin{equation}
W(L,t) = {\left[ { \left< { { {1\over{L^d}}
\sum_i{ {\left( h_i - \overline{h}\right) }^2 } } } \right> } \right] }^{1/2} 
\label{defw}
\end{equation}
for deposition in a $d$-dimensional substrate of length $L$ ($h_i$ is
the height of column $i$ at time $t$, the bar in $\overline{h}$ denotes a
spatial average and the angular brackets denote a configurational average).
For short times it scales as $W\sim t^{\beta}$ and for long times, in the
steady state regime, it saturates at $W_{sat}\sim L^{\alpha}$. The dynamical
exponent $z=\alpha /\beta$ characterizes the crossover from the growth regime
to the steady state regime. For systems belonging to the $EW$ universality
class, we have $\alpha_0= 1/2$, $\beta_0 = 1/4$
and $z_0 = 2$ in $d=1$ (in this paper, the subscript $0$ will refer to
exponents of the $EW$ theory). For systems in the $KPZ$ class, in $d=1$, we
have $\alpha=1/2$, $\beta=1/3$, $z = 3/2$~\cite{Stanley,KPZ,EW}.

Considering the crossover from $EW$ to $KPZ$ scaling in $d=1$, Grossmmann,
Guo and Grant ($GGG$) \cite{GGG} and Nattermann and Tang ($NT$) \cite{NT} (see
also the review by Forrest and Toral \cite{Forrest}) proposed  multiscaling
relations that are equivalent to
\begin{equation}
W(L,t) = L^{\alpha} f ( \frac{t}{t_c}, \frac{L}{\xi_c} ) ,
\label{multiscaling}
\end{equation}
in which $\xi_c \sim t_c^{1/z_0}$. $GGG$ also proposed
that the characteristic time of crossover from $EW$ to $KPZ$ dynamics was
\begin{equation} t_c \sim \lambda^{-\phi} , \label{deffi}
\end{equation}
with $\phi > 0$, since the $EW$-$KPZ$ crossover disappears for $\lambda =0$.
Through  scaling arguments, those authors obtained    
$\phi = z_0 / (\alpha_0 + z_0 - 2 )$, which gives $\phi= 4$ in $d=1$.
This was confirmed through one-loop renormalization group
calculations by $NT$. The scaling analysis of the $KPZ$ equation by Amar and
Family ($AF$)\cite{AF} and the assumption of Family-Vicsek scaling\cite{FV}
were used to show that, in the nonlinear and saturation regimes,
\begin{equation}
W(L,t) \sim L^{1/2}  g( |\lambda| \frac{t}{L^{3/2}}  ) ,
\label{fvkpz}
\end{equation}
in which $g$ is a scaling function and the dependence of $W$ on the parameters
$\nu$ and $D$ of Eq. (\ref{kpz}) was omitted. A generalized scaling
relation equivalent to Eq. (\ref{multiscaling}), which is a more general result
than Eq. (\ref{fvkpz}), was also obtained by Derrida and Mallick in the context
of the connection to the one-dimensional asymmetric exclusion model
\cite{derrida}. Amar and Family \cite{AF} have shown that the scaling form
(\ref{fvkpz}) also predicts a crossover exponent $\phi = 4$.

On the other hand, all previous numerical results suggested
$\phi\approx 3$; for instance, $GGG$ obtained this value using data collapse
methods \cite{GGG}. Thus, it would be desirable to confirm numerically the
scaling properties predicted for a $KPZ$ system in order to solve this
controversy.

The purpose of this work is to study a competitive
growth process with $EW$ to $KPZ$ crossover, involving ballistic deposition
($BD$)~\cite{Stanley,vold}  and  random deposition with surface relaxation
(Family model)~\cite{Family} in $d=1$. In this model, incident
particles aggregate to the deposit according to the rules of $BD$ with
probability $p$ and according to the rules of the Family model with
probability $1-p$. It is known that the Family model is in the $EW$
universality class, while $BD$ is in the $KPZ$ class. This competitive model
was introduced by Pellegrini and Jullien \cite{Pellegrini}, whose main
interest was the connection to the roughening transition present in higher
dimensions. Although it is expected that this model is in the $KPZ$ class for
any $p>0$, the crossover in $d=1$ was not studied in detail in
their original work and, for $p\lesssim 0.3$, effective exponents very near
the $EW$ values were obtained~\cite{Pellegrini}.

Here we will simulate that model in order to analyze the interface width
scaling in the nonlinear regime, the crossover to the saturation regime
and to connect the parameter $p$ and the coefficient $\lambda$ of
the $KPZ$ equation in the corresponding continuum limit. The amplitudes of
typical saturation times and of interface width scaling in the
growth regime are consistent with multiscaling concepts~\cite{GGG,NT,AF} and
refine previous numerical estimates for related systems. The crossover
exponent $\phi =4$ follows directly from our numerical results and, together
with the observed relation $\lambda\sim p^{2.1}$, indicate that the crossover
at small $p$ is very slow. The analysis of this apparently simple problem
shows that, in order to obtain reliable asymptotic exponents governing various
quantities, it is essential to account for corrections to the leading terms
in the scaling relations. Thus, this work may also be relevant to the
analysis of other systems with slow crossover to $KPZ$ scaling, whose interest
increased after the recent debate on the problem of Fisher waves and their
discrete realizations in $d=1$ dimensions~\cite{doering,tripathy,blythe,moro}.
For that reason, the crossover effects identified in our simulations' data
will be discussed in detail.

It is also relevant to point out that a related competitive model
was recently studied in $d=1$ and $d=2$ \cite{tales1}, showing evidence
of the asymptotic $KPZ$ behavior. However, that work did not study the
relation between the parameters of the discrete and the continuous ($KPZ$)
model nor the scaling amplitudes that will be considered here.

The rest of this work is organized as follows.
In Sec. II, we will define precisely the discrete model and 
connect it to the $KPZ$ equation using the scaling properties of
the growth velocity. In Secs. III and IV we will present results for the
interface width scaling in the discrete model at the nonlinear growth regime
and at the steady state regime, respectively. In Sec. V we summarize our
results and present our conclusions. 

\section{The discrete model and its connection to the $KPZ$ theory}
\label{secconnection}

We considered a model in which particles are 
aggregated following the rules of $BD$  with  probability
$p$ or the rules of random deposition with surface relaxation (Family
model) with probability $1-p$. In $BD$ (Fig. 1a), the incident particle
follows a straight trajectory perpendicular to the surface and sticks upon
first contact with a nearest neighbor occupied site. It leads to the formation
of a porous deposit. In the Family model (Fig. 1b), the particle falls towards
the surface along the incident column and sticks at the top of that column if
its height is lower than or equal to the heights of the neighboring columns.
Otherwise, the particle diffuses to the neighboring column which has the lowest
height and, if two or more neighbors have the same height, it chooses
one of them randomly. 

For $p=0$, we have the Family model, which is in the $EW$ universality
class. For any $p \neq 0$, in $d=1$, we expect the $BD$ process to
change the universality class to $KPZ$ in the continuum limit (see the analysis
in Ref. \cite{YKS} for a related model). Then the
coefficient $\lambda$ of the nonlinear term vanishes with $p$ in
the form
\begin{equation}
\lambda \sim p^\gamma ,
\label{connection}
\end{equation}
with $\gamma >0$ (to be estimated below).
For small $p$ and sufficiently large $L$, the interface width
$W(L,t,p)$ must scale analogously to the weak coupling regime of the KPZ theory
\cite{Forrest,Pellegrini}, in which three regimes were identified: a
linear ($EW$) growth  regime at early times ($ t \ll t_c$), a nonlinear ($KPZ$)
growth regime for  $t_c \ll  t \ll \tau$  and the saturation regime for
$t\gg \tau $, as illustrated in Fig. 2 ($\tau$ is the characteristic
time for the interface width saturation).

In order to calculate the exponent $\gamma$, we considered the
scaling of the interface growth velocity. The difference between the growth
velocity in an infinitely large substrate, $v_{\infty}$, and the velocity in
the steady state of a finite lattice (thick films), $v(L)$, scales
as~\cite{Krug,Krug1}
\begin{equation}
\Delta v(L) \equiv v_{\infty} -v(L) \sim \lambda L^{-\alpha_{\Vert} } ,
\label{deltav}
\end{equation}
with  $\alpha_{\Vert} = 1$ in $d=1$ \cite{Krug}. Defining
\begin{equation}
b_v(L) \equiv \Delta v(L) \times L,
\label{bv}
\end{equation}
we expect that, as $L\to\infty$,
\begin{equation}
b_v(L) \to B_v = B\lambda ,
\label{bvinf}
\end{equation}
where $B$ is a constant.

In the discrete model, $\lambda$ varies with $p$, consequently $b_v$ is a
function of $L$ and $p$ which has a finite limiting value $B_v(p)$ as
$L\to\infty$. For very large $L$, Eqs. (\ref{connection}) and (\ref{bvinf})
shows that $B_v$ scales with $p$ with exponent $\gamma$.

Simulations of the model were performed in lattice of lengths
from $L=16$ to $L=4096$ until the saturation regime, and in lattices
with $L=2^{16} = 65536$ during the growth regime (linear and nonlinear), for
several values of the probability $p$ between $p=0.15$ and $p=0.5$. The results
presented in this paper are averages typically over ${10}^5$ realizations for
the smallest lattices ($L\leq 256$), ${10}^4$ realizations for $256\leq L\leq
4096$ and ${10}^2$ realizations for $L=65536$.
The growth velocities were calculated from numerical derivatives
of the average heights of the deposits, with accuracies from $5$ to $6$ decimal
places, in lattices of lengths $L\leq 128$ ($L\leq 512$ for $p=0.15$).
We considered the data for $L=65536$ as representative of an infinite lattice
in the growth regime (some simulations in $L=131072$ supported this
assumption), and also obtained $v_{\infty}$ with high accuracy.
These data provided estimates of $b_v(p,L)$ with accuracy from $0.5\%$ to
$5\%$. For larger lengths, poorer results were obtained due to the much smaller
number of realizations. 

In Fig. 3 we show $b_v(p,L)$ versus $1/L$ for the three smallest values of $p$
considered in this work. The variable $1/L$ in the abscissa was the best choice
to represent finite-size corrections in $b_v$ as $L\to\infty$, and is
related to higher order terms ($1/L^2$) in Eq. (\ref{deltav}). Such
scaling corrections have been previously observed in the analysis of small
$L$ data for $BD$ and for the restricted solid-on-solid model by Krug and
Meakin \cite{Krug}. The corrections were considered in the extrapolation of the
data in Fig. 3, which provided estimates of $B_v(p)$ for several values of $p$
(intercepts with the vertical axis in Fig. 3).

Crossover effects may be crucial in the extrapolation procedure discussed
above, and may severely affect the estimates of $B_v$ for small $p$. For
$0.25\leq p\leq 0.5$, four values of $b_v$ ($16\leq L\leq 128$) were
well fitted by straight lines in the $b_v(p,L)\times 1/L$ plots (these
results were not shown in Fig. 3, except for $p=0.25$). For
$p=0.2$, the data for $32\leq L\leq 256$ confirm the presence of the $1/L$
correction and was also used to estimate $B_v$ (the estimate for $L=16$
deviates from this trend). On the other hand, for $p=0.15$, the result for
$L=256$ showed a crossover in $b_v$, which suggested calculations
for $L=512$. Fig. 3 shows that $b_v(0.15,L)$ slowly increases for $16\leq
L\leq 128$, but decreases for $128\leq L\leq 512$. Consequently,
the extrapolation considered only the three last points (see Fig. 3) and 
gave $B_v\approx 0.057$. However, if the extrapolation to
$L\to\infty$ was performed only with results for $L\leq 128$, then a $7\%$
larger value of $B_v$ would be obtained. Smaller values of $p$ were not
studied here because such crossover would appear for much larger $L$ and,
consequently, the extrapolations based on small systems' data would provide
unreliable estimates of $b_v(p,\infty)$. 

In Fig. 4 we show $\ln{\left[ B_v(p)\right] }$ versus $\ln{p}$ using the
extrapolated values of $B_v$, as discussed above. The linear
fit in Fig. 4 gives $B_v \sim p^{2.1}$. Considering the error bars in $B_v$,
we obtain an exponent $\gamma = 2.1\pm 0.2$ (Eq. \ref{connection}).

The large value of the exponent $\gamma$ explains the crossover effect
discussed above. Since $\lambda$ decreases rapidly with $p$, the coefficient of
the leading term in $\Delta v$ (Eq. \ref{deltav}) is small compared to higher
order corrections ($1/L^2$, $1/L^3$ etc) for small $p$. Thus, very large
values of $L$ are needed to provide reliable extrapolations with a single
correction term, which prevented us to use values of $p<0.15$ in our study.

\section{Interface width at the nonlinear growth regime}
\label{roughness1}

In the nonlinear growth regime ($t_c \ll t \ll \lambda^{-1} L^z$) for
sufficiently large substrates in $d=1$, the interface width does not depend on
$L$ (weak finite-size effects). Then the scaling function of Eq. (\ref{fvkpz})
behaves as
\begin{equation} g(x) \approx C x^{\beta} \;\; , \;\; \beta = 1/3 \;\; ,
\label{growthg}
\end{equation}
with constant $C$, so that $W$
does not depend on $L$, except for vanishing corrections to scaling.
Consequently, the $\lambda$-dependent scaling of $W$ in this regime is
\begin{equation}
W \approx C \lambda^{\beta} t^{\beta} .
\label{defC}
\end{equation}
In this section, we will verify this $\lambda$-dependence through a careful
analysis of simulations' data of our discrete model.

However, first we will show that Eq. (\ref{defC}) gives $\phi =4$ (Eq.
\ref{deffi}) in a simple way, as follows. The crossover $EW-KPZ$ (at $t\sim
t_c$) occurs when the scaling relation (\ref{defC}) matches the $EW$ scaling
\begin{equation}
W(t,L) \approx  C'  t^{\beta_0}    ,
\label{defbeta0}
\end{equation}
with $C'$ constant. Thus
\begin{equation}
{t_c}^{\beta_0}  \sim  {\lambda}^{\beta} {t_c}^{\beta} ,
\label{match}
\end{equation}
then we obtain $\phi = {\beta \over{(\beta - \beta_0 )}} =
{z_0 \over {(z_0-z)}} = 4$.

We conclude that the numerical test of Eq. (\ref{defC}), in particular of the
dependence on the parameter $\lambda$, may be used to test the proposal $\phi
=4$. The first step is to extract the amplitude of $t^\beta$ scaling in Eq.
(\ref{defC}), which motivates the definition of the amplitude $a(p,t)$ as
\begin{equation}
a(p,t) \equiv W(L\to\infty ,t)/t^{1/3} .
\label{defapt}
\end{equation}
In Fig. 5a we show $a(p,t)$ versus $1/t^{1/3}$ for several values
of $p$, using the data for $L=2^{16}=65536$. Different variables in the form
$1/t^x$ ($x>0$) were tested in the abscissa, but the variable
$1/t^{1/3}$ of Fig. 5a provided the best linear fits for most values of $p$.
The fact that in Fig. 5a $a(p,t)$ is still decreasing for large $t$ indicates
the presence of a constant (independent of $t$) correction to the leading
behavior in Eq. (\ref{defC}). It proves again the relevance of accounting for
scaling corrections in this problem, although we are not able to justify these
corrections on theoretical grounds.

As $t\to\infty$, $a(p,t)$ converges to a finite limiting value
\begin{equation}
A(p) \equiv a(p,\infty) .
\label{defA}
\end{equation}
$A(p)$ is the complete amplitude of $t^{\beta}$ scaling of the interface
width in the nonlinear growth regime (Eq. \ref{defC}). Our estimates of
$A(p)$ were obtained from linear extrapolations of $a(p,t)\times 1/t^{1/3}$
plots to $t\to\infty$ (intercepts with the vertical axis in Fig. 5a).

From Eq. (\ref{defC}), it is expected that the amplitude $A(p)$
scales as $\lambda^\beta$. From the connection relation (\ref{connection}), it
is expected that
\begin{equation}
A(p) \sim p^{\delta} ,
\label{defdelta}
\end{equation}
with
\begin{equation}
\delta = \gamma \beta .
\label{reldelta}
\end{equation}
Then, the test of Eq. (\ref{defC}) reduces to the test of Eq. (\ref{reldelta})
for the amplitude exponent $\delta$.

In order to calculate the exponent $\delta$ in the relation (\ref{defdelta}),
our first step was to plot $\log{\left[ A(p)\right]}$ versus $\log{p}$, but we
noticed that it showed decreasing slopes as $p$ decreased. We
analyzed the evolution of the slopes of $\log{\left[ A(p)\right]}\times\log{p}$
plots by calculating the following effective exponents for consecutive values 
$p=p'$ and $p=p''$:
\begin{equation}
\delta_p = { {\ln\left[ a(p',\infty)/a(p'',\infty) \right]} \over
{\ln\left( p'/p'' \right) } } \;\; , \;\; p = \sqrt{p'p''} ,
\label {defdeltap}
\end{equation}
so that, as $p\to 0$ ($\lambda\to 0$), we expect that
$\delta_p \to \delta$.

In Fig. 5b we show $\delta_p$ versus $p^2$, which gives a reasonable linear
fit and indicates that $\delta = 0.7\pm 0.2$. Again the variable
$p^2$ in the abscissa is the one that provides the best linear fit of the
central estimates of $\delta_p$, chosen among other variables in the form $p^y$
($y>0$). In Fig. 5b, the effective exponents systematically decrease as
$p$ decreases, which reflects our previous observation of decreasing slopes in
$\log{\left[ A(p)\right]}\times\log{p}$ plots.

Our estimates $\delta = 0.7\pm 0.2$ and $\gamma =2.1 \pm 0.2$ (Sec.
\ref{secconnection}) are consistent with relation (\ref{reldelta}) with
$\beta = 1/3$. Even considering that the error bars are large, it is relevant
to notice that the central estimates confirm that relation exactly, which
gives additional support to our analysis. 

\section{Interface width near and at the steady state regime}
\label{roughness2}

Our numerical results in the steady state regime provide additional support for
the scaling picture proposed for the problem.

From Eq. (\ref{fvkpz}), we
expect that the crossover from the nonlinear to the steady state
regime takes place at a characteristic time $\tau$ that scales as
\begin{equation}
\tau \sim \lambda^{-1} L^z ,
\label{deftau}
\end{equation}
with $z=3/2$ in $d=1$. In this section, our main purpose is to test the
$\lambda$-dependence of this characteristic time in our discrete model.

The saturation time $\tau$ is usually estimated using some arbitrary recipe.
Here, instead of estimating the saturation time $\tau$ (which may be defined
from the time dependence of the interface width as it converges to the
saturation value), we calculated a characteristic time $\tau_0$ which is
proportional to $\tau$, according to a recently proposed method~\cite{Reis}.
That method provided accurate estimates of dynamic exponents for several
growth models in $d=1$ and $d=2$, including the Family and the $BD$ models.

First, the saturation width $W_s$ is
estimated, for fixed $p$ and $L$. Then we define $\tau_0$ through
\begin{equation}
W\left( \tau_0\right) = kW_s ,
\label{deftau0}
\end{equation}
with fixed $k$ ($k\lesssim 1$) \cite{Reis}.
Using the Family-Vicsek relation $W(L,t) = L^{\alpha} f(tL^{-z})$ and
considering that $W_s\sim L^{\alpha}$, we conclude that $\tau_0\sim L^z$,
i.e., $\tau_0$ is proportional to the saturation time $\tau$. For the
particular case of a $KPZ$ system, Eq. (\ref{fvkpz}) gives
\begin{equation}
\tau_0 \sim \lambda^{-1} L^{3/2} .
\label{tau}
\end{equation}

Extending the procedure of previous work~\cite{Reis}, we
considered $k =1-1/e = 0.6321\dots$ in Eq. (\ref{deftau0}) to estimate
$\tau_0$. This value of $k$ gave $\tau_0\approx \tau$ for $BD$, where $\tau$
was estimated from the decay of $W_s-W$~\cite{Reis}. In the present model, for
fixed $p$, we calculated the ratios $\tau_0/L^{3/2}$ for several lengths $L$
and obtained the asymptotic amplitude
\begin{equation}
D(p) = \frac{\tau_0}{L^{3/2}} \;\; ,
\;\; L \to \infty .
\label{defD}
\end{equation}
The extrapolation procedure follows the same lines of the calculation of
$A(p)$ from $a(p,t)$ in Sec. \ref{roughness1}. However, only results for $p\geq
0.2$ could be obtained using data for lattice sizes $L\leq 4096$, since the
saturation for smaller values of $p$ is typically of $EW$ type ($\tau\sim
L^2$) in this range of $L$.

From Eqs. (\ref{defD}) and (\ref{tau}), we expect that
$D(p)\sim {\lambda}^{-1}$. Consequently, it must scale with $p$ as
\begin{equation}
D(p) \sim p^{-\gamma} .
\label{scalingD}
\end{equation}
In Fig. 6 we show $\ln{D(p)}$ versus $\ln{p}$, with a linear fit that gives
$D(p) \sim p^{-2.1}$. This result is consistent with the independent
estimate of $\gamma$ from Eq. (\ref{connection}) (Sec. \ref{secconnection}).

We also analyzed the scaling of the saturation width $W_s$. 
For lattice sizes $L \leq 1024$, we obtained $ W_s \sim L^{\alpha}$
with $\alpha=1/2$ and weak corrections to scaling.
Using the data for $L=1024$, we defined
\begin{equation}
{\Delta W}_s \equiv W_s(p) -W_s(0)
\label{deltaws}
\end{equation}
as the difference between the saturation width for a given probability $p$ and
the saturation width for the Family model ($p=0$).

In Fig. 7 we show $\ln{{\Delta W}_s}$ versus $\ln{p}$. The linear fit suggests
${\Delta W}_s \sim p^{3/2}$, thus we obtain the complete form for the
saturation width as
\begin{equation}
W_s \approx ( C_1 + C_2 p^{3/2}) L^{\alpha}
\label{scalingws}
\end{equation}  
with $\alpha=1/2$, $C_1$ and $C_2$ constants.
The amplitude of $W_s$ scaling is
${\left( D\over{24\nu}\right)}^{1/2}$~\cite{Krug1}, i.e., the heights'
fluctuations depend only on the parameters $\nu$ and $D$ of the
$KPZ$ equation (\ref{kpz}), but not on the nonlinearity parameter $\lambda$.
Thus we conclude that the dependence on $p$ in Eq. (\ref{scalingws}) is
related to the dependence on $p$ of the surface tension parameter $\nu$: when
$p$ decreases, the amplitude in Eq. (\ref{scalingws}) decreases, then
the parameter $\nu$ increases. Indeed, this term is physically expected to
increase in the crossover from $BD$ (low $\nu$) to the Family model (high
$\nu$).

\section{Summary and conclusions}
\label{conclusions}

We studied a competitive growth model in $1+1$ dimensions involving two
dynamics: ballistic deposition  with probability $p$ and random deposition
with surface  relaxation (Family model) with probability $1-p$. This 
model is  a discrete realization of the continuum $KPZ$ equation with an 
adjustable nonlinear coupling $\lambda$ related to $p$.
At the critical probability $p_c=0$, the process belongs to the  $EW$ 
universality class, while any finite value of $p$ drives the system to $KPZ$ 
class. 

We established the connection  between the parameters $p$ and $\lambda$ as
$\lambda\sim p^{2.1}$ and showed that $W\sim p^{0.7}t^{1/3}$ in the growth
regime. This indicates that the discrete model presents a  very slow 
crossover from $EW$ to  $KPZ$  scaling at small  values of $p$, since the
crossover time is $t_c\sim {\lambda}^{-4} \sim p^{-8.4}$. This slow crossover
explains the discrepancies in the  effective exponents $\beta$ measured in
that regime in previous works \cite{Pellegrini}.

We also obtained the saturation time $\tau \sim p^{-2.1} L^{3/2}$. 
The condition $\tau\gg t_c$ is necessary to observe the crossover to $KPZ$
scaling, while the opposite condition leads to $EW$ saturation without an
intermediate $KPZ$ growth of the interface width. A critical system
size $\xi_c$ separates systems which present $EW$ or $KPZ$ saturation, and
$\xi_c$ can be estimated from the condition $\tau\sim t_c$, which gives
$\xi_c \sim \lambda^{-2} \sim p^{-4.2}$. This large exponent proves that
simulations in very large system sizes are necessary in order to observe all
features of $KPZ$ scaling for small $p$.

Our results are consistent with the scaling theories for the weak
coupling regime of the $KPZ$ equation proposed by several authors and refine
previous numerical analysis. Then we expect that the methods presented here may
be helpful to analyze other growth models with slow crossovers to $KPZ$
scaling, in which scaling theories cannot be easily developed.

%~~~~~~~~~~~~~~~~~~~~~~~~~~~~~~~~~~~~~~~~~~~~~~~~~~~~~~~~~~~~~~~~~~~~~~~~~~~
%~~~~~~~~~~~~~~~~~~~  REFERENCES  ~~~~~~~~~~~~~~~~~~~~~~~~~~~~~~~~~~~~~~~~~~
%~~~~~~~~~~~~~~~~~~~~~~~~~~~~~~~~~~~~~~~~~~~~~~~~~~~~~~~~~~~~~~~~~~~~~~~~~~~

\vskip1cm
 
\begin{figure}
\caption{ (a) The aggregation rules of ballistic deposition, in which the
sticking position of each incident particle is marked with a cross. (b) The
aggregation rules of the Family model, in which the relaxation of incident
particles to their sticking positions is indicated by arrows. The incident
particle at the right has equal probabilities to stick at any one of the
neighboring columns.}
\label{fig1}
\end{figure}

\begin{figure}
\caption{  For small values of $p$  and sufficiently  large
$L$, the interface width $W(t,L,p)$ presents three regimes:  
a linear ($EW$) growth  regime at early times ($ t \ll t_c$), 
a nonlinear ($KPZ$) growth regime for  $t_c \ll  t \ll \tau$  and 
the saturation regime for $t\gg \tau $. 
 }
\label{fig2}
\end{figure}
 
\begin{figure}
\caption{     
 $b_v(p,L)  \equiv \Delta v \times L  $ versus
$1/L$ for $p=0.25$ (squares), $p=0.2$ (triangles) 
and $p=0.15$ (crosses). Solid lines are least squares fits of the data for
larger $L$.}
\label{fig3}
\end{figure}  
 
\begin{figure}
\caption{  Log-log plot of  
$B_v(p)$ as a function of $p$. The linear fit gives an exponent $\gamma =
2.1\pm 0.2$.  }
\label{fig4}
\end{figure}       
                        
\begin{figure}
\caption{ 
(a) Interface width amplitude in the nonlinear regime
$a(p,t)\equiv W\left( L\to\infty ,t\right) /t^{1/3}$ as a function  of  
$1/t^{1/3}$ for $p=0.4$,
$p=0.3$, $p=0.25$, $p=0.2$ and   $p=0.15$ (from top to bottom); 
(b)  Effective exponent $\delta_p$ versus $p^2$, with a linear fit that gives
the exponent $\delta\approx 0.7$ as $p\to 0$.  }
\label{fig5}
\end{figure}

\begin{figure}
\caption{ Log-log plot of the amplitude  
$ D(p)$ ($\tau_0 /L^{3/2}$ as $L\to\infty$) as a function of $p$. The linear
fit gives  $D(p) \sim p^{-\omega}$ with $\omega\approx 2.1$ . 
}
\label{fig6}
\end{figure}

\begin{figure}
\caption{  Log-log plot of  
${\Delta W}_s \equiv W_s(p) -W_s(0)$ as a function of $p$, using  data 
for $L=1024$. The linear fit suggests ${\Delta W}_s \sim p^{3/2}$.
}
\label{fig7}
\end{figure}

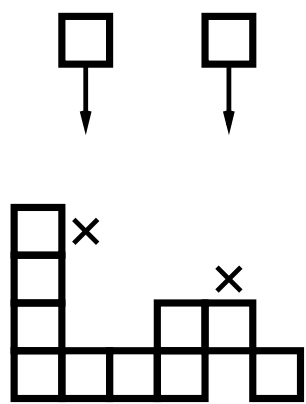
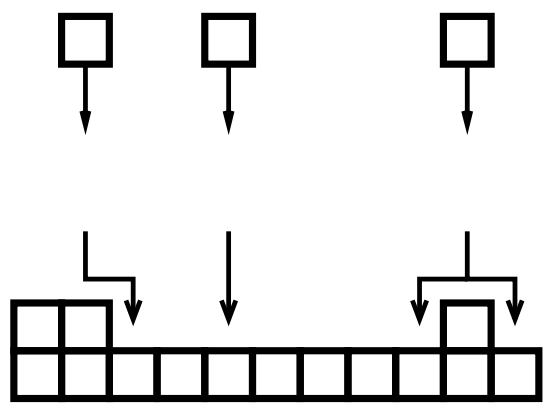
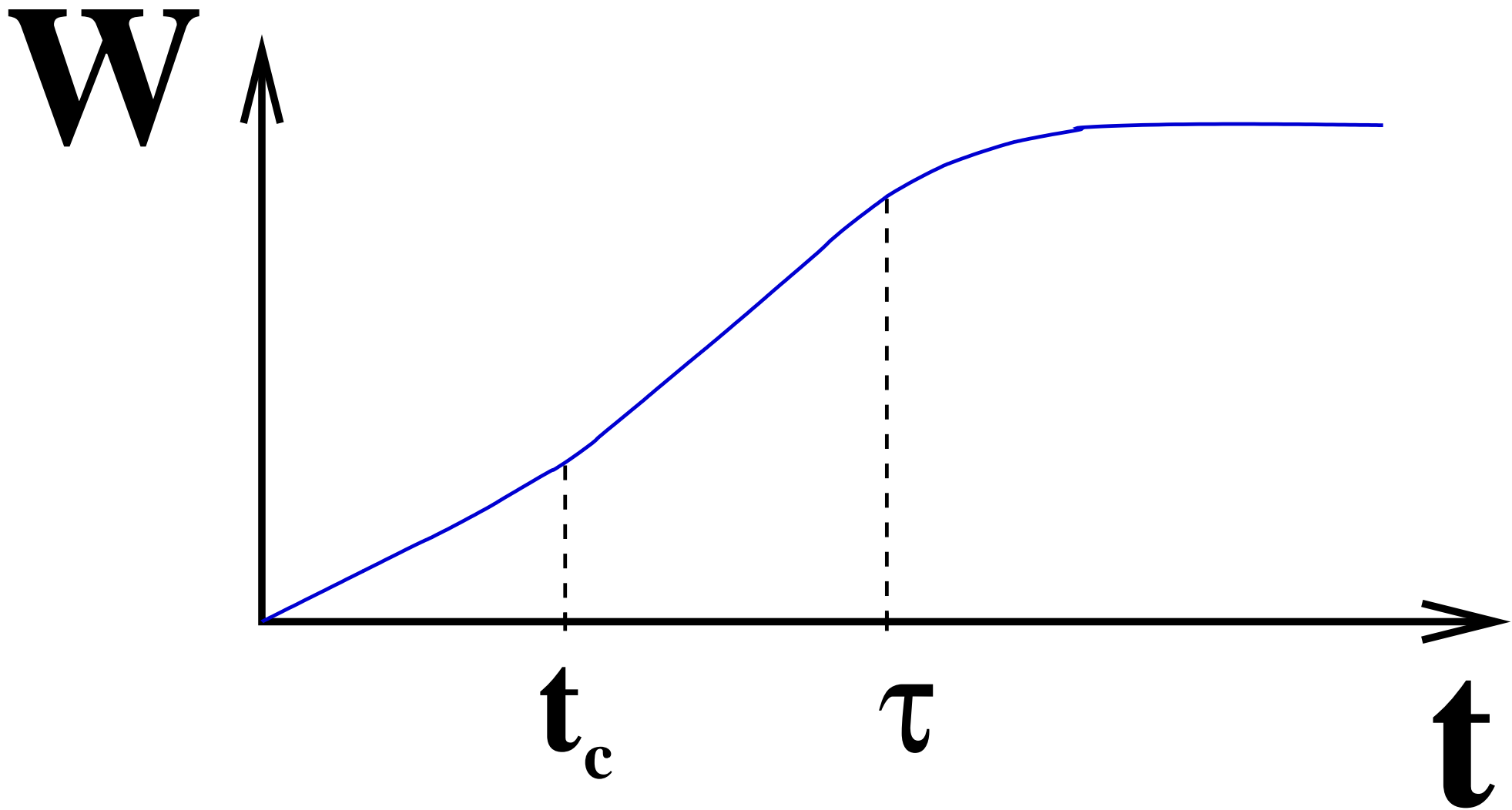
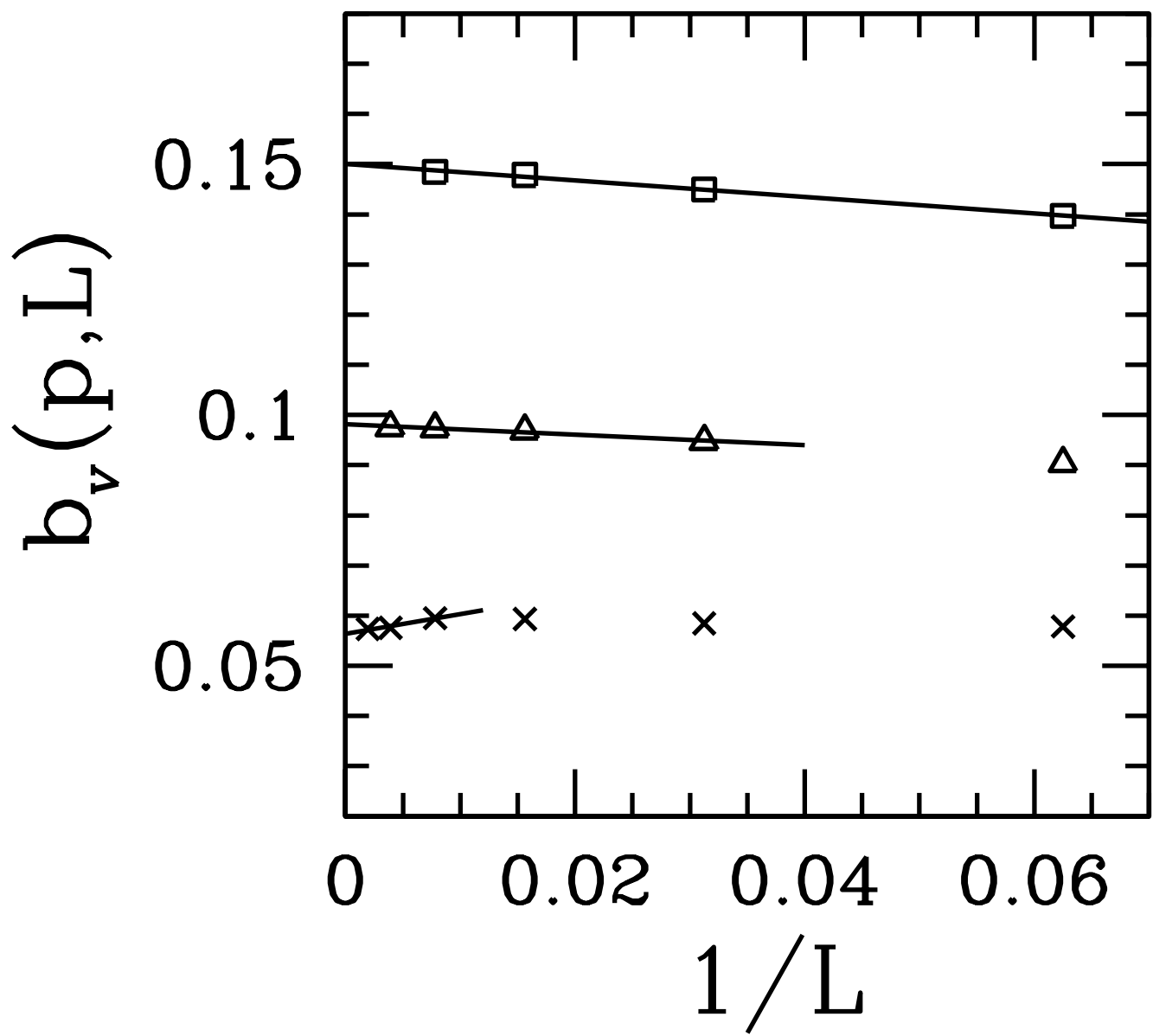
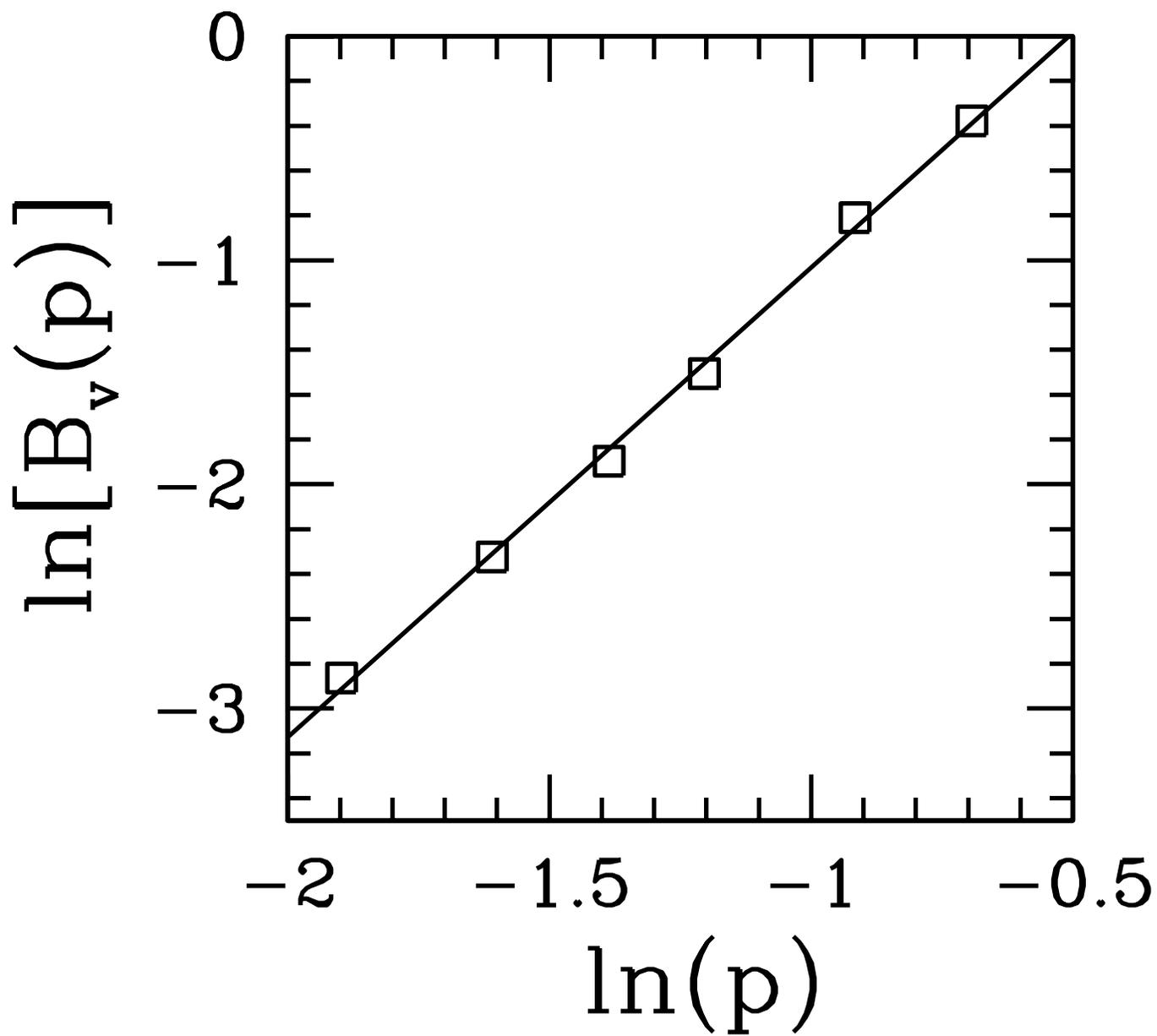
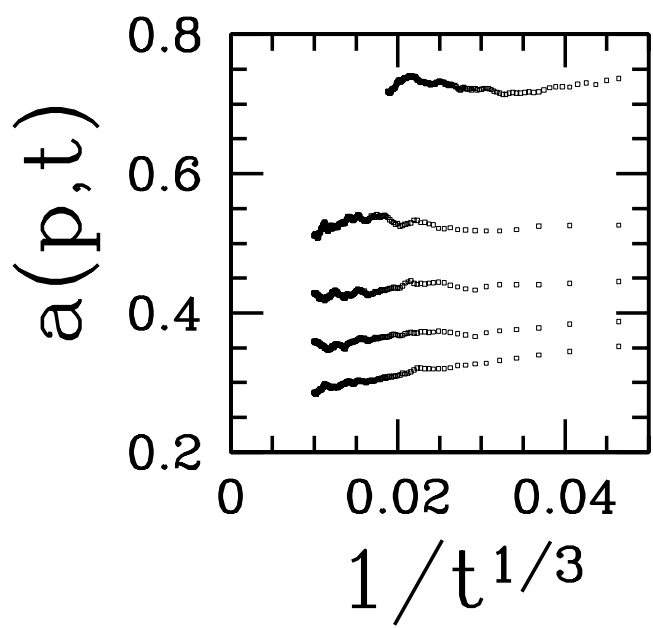
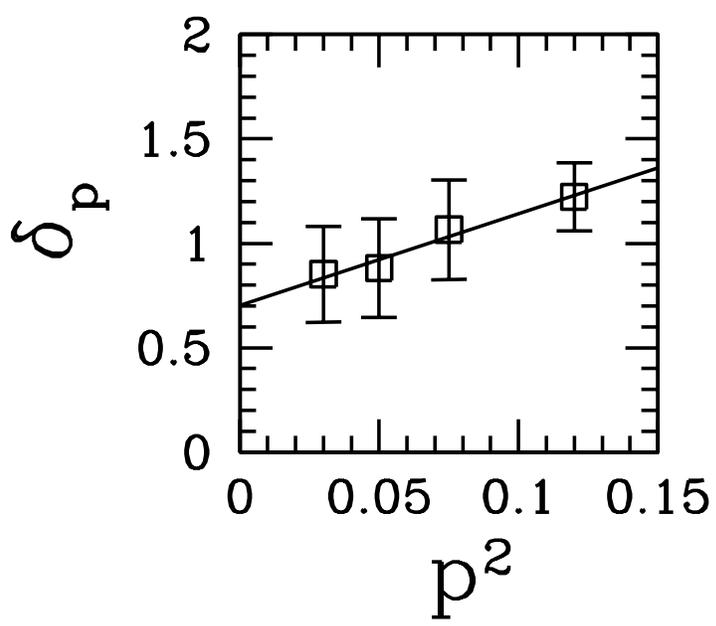
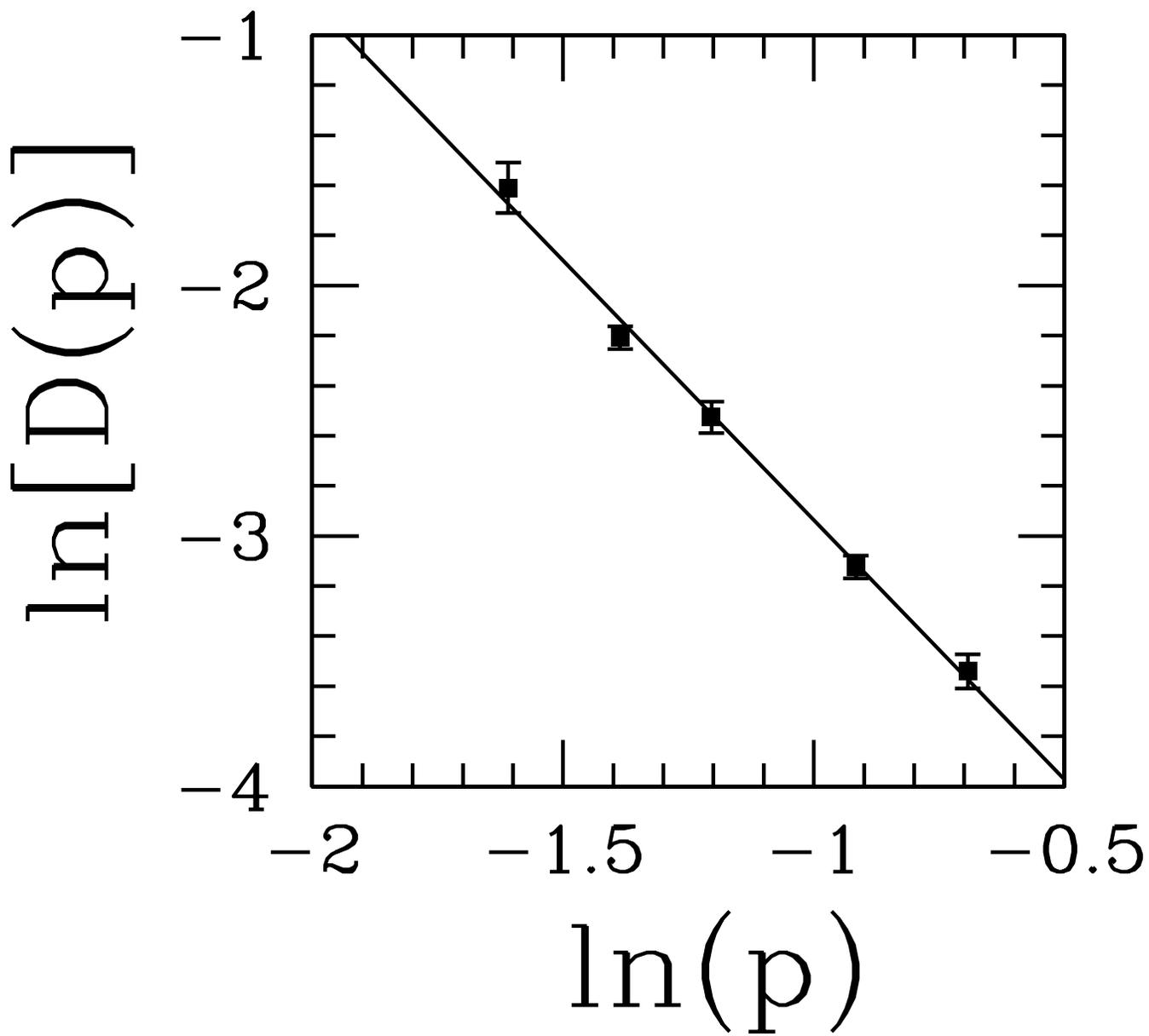
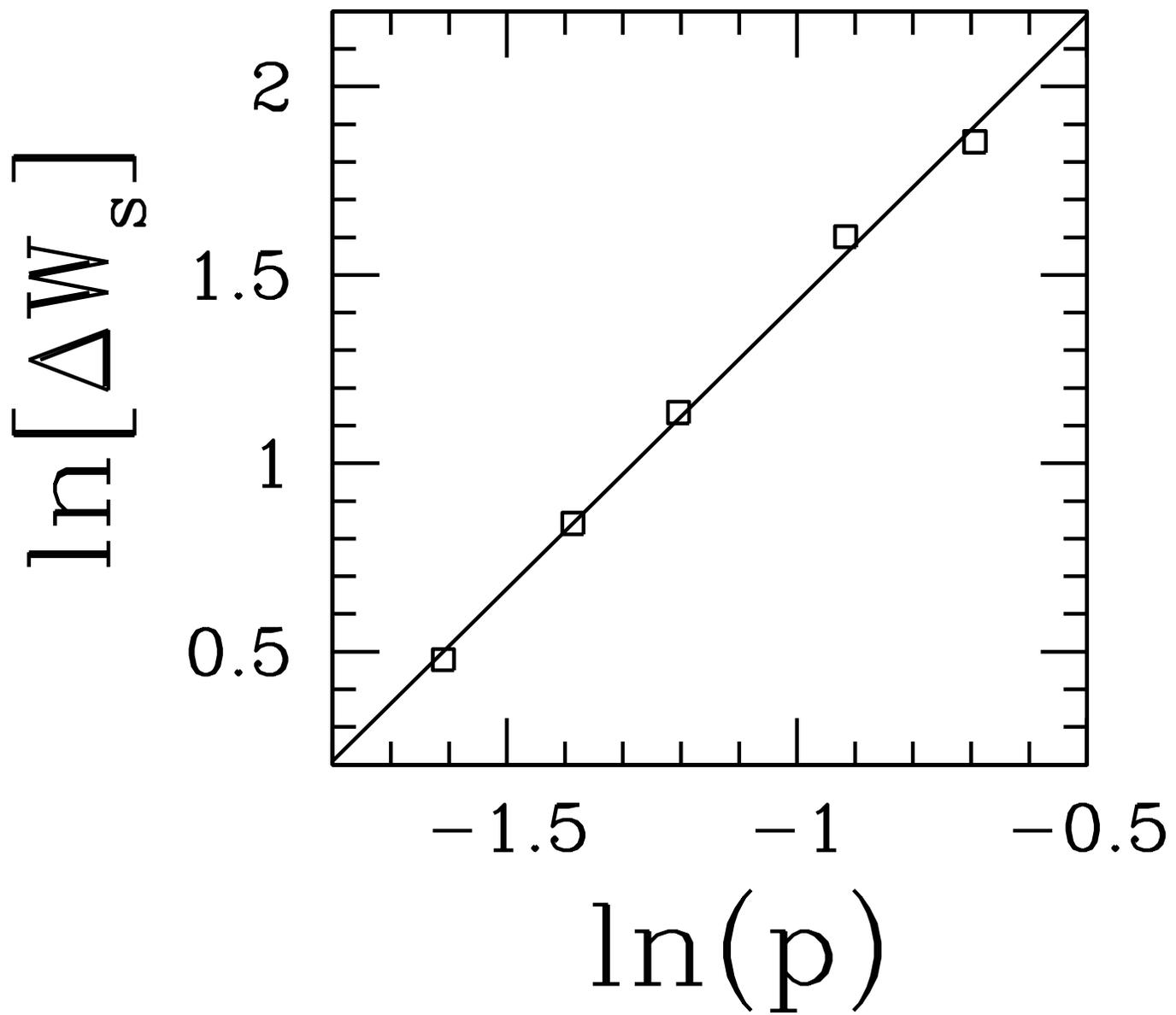
\end{document}